# Tunable Polymer/Air Bragg Optical Microcavity Configurations for Controllable Light-Matter Interaction Scenarios

Chirag Chandrakant Palekar[1], Arash Rahimi-Iman[1]

[1]Faculty of Physics and Materials Sciences Center, Philipps-Universität Marburg, D-35032, Marburg, Germany.

**Abstract**

**Complex optical systems such as high-quality microcavities enabled by advanced lithography and processing techniques paved the way to various light-matter interactions (LMI) studies. Without lattice-matching constraints in epitaxy, coating techniques or shaky open cavity constructions, sub-micrometer-precise lithographic development of a polymer photoresist paves the way to polymer microcavity structures for various spectral regions based on the material's transparency and the geometrical sizes. We introduce a new approach based on 3D nanowriting in photoresist, which can be employed to achieve microscopic photonic Fabry-Pérot cavity structures with mechanically-tunable resonator modes and polymer/air Bragg mirrors, directly on a chip or device substrate. We demonstrate by transfer-matrix calculations and computer-assisted modelling that open microcavities with up to two "air-Bragg" reflectors comprising alternating polymer/air mirror-pair layers enable compression-induced mode tuning that can benefit many LMI experiments, such as with 2D materials, nanoparticles and molecules.**

## Introduction

Optical microcavities play an important role in the investigation of a wide range of research areas, such as light-matter interaction (LMI) [1][2][3][4], nonlinear optics [5][6], and quantum information processing [1][7][8]. Various high-quality microcavities [1][9][10][11] have been explored for decades and enabled the hunt for ultralow-threshold nanolasers [5][12][13], opened up fundamental cavity-QED experiments [14][15][16], and the study of Bose-Einstein-like condensation (BEC) of polaritons in solids [17][18][19][20]. Microcavities have reached popularity not only for conventional (more energy-efficient) lasers, but also for polariton physics [3][21], as well as the novel field of polariton chemistry [22], and became indispensable for optical quantum technologies [23][24][25]. In fact, the list of abundant research directions with confined light fields cannot be projected adequately in this short summary here.

Recently, a variety of new and practical configurations of optical microresonators were used for the investigation of LMI with ultrathin van-der-Waals materials, such as two-dimensional (2D) transition-metal dichalcogenides (TMDC), colloidal quantum dots, fluorescent dyes, III-V semiconductors and many other materials [3][11][26]. For the investigation of LMI, be it (quantum) optoelectronic or optomechanic coupling, most commonly used microcavities are (monolithic) planar/ Fabry-Perot (FP) microcavities [14][27], open tunable fiber-based concave mirrors microcavities [28][29], (total-internal reflecting) whispering gallery modes [30][31], (low-mode-volume) photonic crystal nanocavities [32][33][34], and plasmonic metal cavities (with possible ohmic losses) [3][35], as well as the air-gap type microresonators [36][37][38]. Among the various designs, those highly versatile, spectrally tunable and relatively simple open-cavity configurations have already covered a large spectrum of applications ranging from cavity quantum electrodynamics (CQED) [7][39] to (fiber-coupled) optoelectronical or optomechanical devices [29][40][41] as well as sensors [42][43][44].

Many different and unique approaches have been already demonstrated to improve the confinement of the light and functionality of cavity systems [1][3]. Some approaches such as photonic crystal membranes and whispering gallery modes do provide significant confinement of light in all the three spatial dimensions, whereas open cavity configurations with an air-gapped microresonator such as fiber-based FP cavities provide intrinsic tunability in both the spectral and spatial domains [10][29][41][42][45] as flexible light-matter interfaces. Often, a top-down nanotechnology approach is used to define precise (monolithic) resonator structures, or movable reflectors are combined to form tunable open resonators. In contrast, 3D (nano-)printing offers on-demand bottom-up production of cavity configurations nearly arbitrarily on various substrates and in combination with different active materials, around quantum emitters, in combination with fluids and gases, and even in a disposable fashion.

The tunable nature of open (air-gap) FP microcavities is a key advantage which provides in-situ control over LMI through longitudinal mode tuning, lateral mode positioning, compatibility with different (nano-)materials and access to the intracavity space. However, open microcavity configurations are typically susceptible to vibrations and unstable at relatively large cavity length. In contrast, nano-printing of FP microcavities around a target material could open up a new path to tailor-made optical resonators. To achieve this, suitable 3D-printable micromirror structures need to be designed and developed for the incorporation of the desired active material prior to or after completion of the cavity structure, as appropriate for the given experiment. While mirror production can easily rely on metallization of surfaces, wavelength tunable on-demand deposition of (top-)mirrors can be best addressed with layered dielectric mirrors based on the distributed Bragg reflector (DBR) concept, which can conveniently be formed by polymer/air layer pairs offering considerable refractive index contrast. In case the rigidity and elasticity allow for pressure-induced thickness changes of layered polymer/air-gap structures, even mechanically tunable optical microcavities can be envisioned. Thus, research platforms for tunable Rabi splitting, BEC studies, polariton chemistry, Purcell-enhanced single photon sources and field-enhancement-benefitting nonlinear optics can arise, considering the estimated reasonably-high Q factors, wavelength or structure design flexibility, and stability of such a printed photonic microstructure.

Here in this work, two promising printed cavity system configurations are discussed utilizing simulations of optical properties of the microcavities with and without active materials based on the transfer matrix method (TMM). The here presented simulation study for different selected layer thicknesses in the polymer/air ("air-Bragg") microcavity, i.e. of air and the polymer material, is supported by additional stress analysis using a finite element analysis (FEA) for the examination of mechanical stability of the designed structures. Thereby, we demonstrate the practicality and conceptual feasibility of mode-tunable strong-coupling experiments with a van-der-Waals semiconductor monolayer incorporated theoretically in the polymer/air microcavity. In the future, laser nanoprinting of (integrated) photonic structures directly on the chip with similar and more advanced polymer optical microreflectors and microresonators promise great flexibility for optoelectronic, optomechanic, nanophotonic and nonlinear-optics applications.

**The Optical Microcavity Configurations**

The optical mirrors in the form of DBRs, i.e. alternating refractive index layers with thicknesses of $i\,\lambda_0/(4\,n_{mat})$, where $\lambda_0$ is the wavelength in vacuum, $n_{mat}$ is the refractive index of the material used and $i$ = 1,3,5 and so forth (odd integer numbers), solely rely on the resist material (and air) and are designed to work well with WS$_2$ monolayer excitonic resonances. In the following, $\lambda = (\lambda_0/\,n_{mat})$.

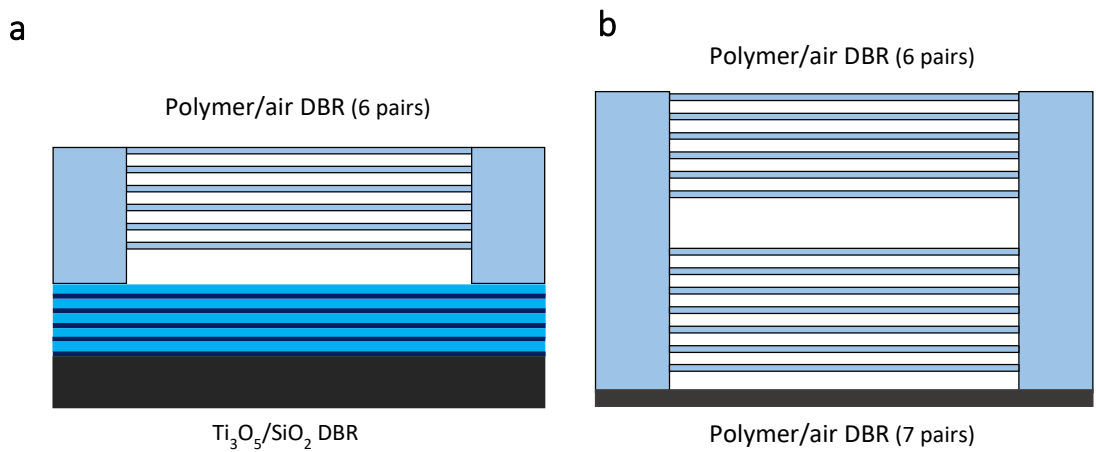

**Figure 1:** *Schematic of the possible microcavity configurations incorporating the printed polymer/air-Bragg reflectors. **a)** "Air-Bragg"-Ti$_3$O$_5$/ SiO$_2$ DBR microcavity. **b)** Polymer/air-Bragg-reflector microcavity.*

The first configuration (**Fig. 1a**) is composed of a conventional dielectric mirror, comprising 6 pairs of $SiO_2$ and $Ti_3O_5$, as the bottom mirror as well as substrate for the active medium, and the polymer/air-Bragg structure as the top mirror (in the following "air-Bragg" reflector). The second configuration (**Fig. 1b**) aims at directly printing two air-Braggs on top of one-another with an appropriate distance (cavity spacer) between them to form a microresonator incorporating the active medium, which needs to be (in most cases) inserted in an intermediate step. For layered materials, this is sketched in the Supplementary Information. Later on, the desired tunability of cavity modes can be introduced by application of mechanical pressure on the polymer/air microstructure. If appropriately constructed (with a stable enough configuration), in principle, one can compress mirrors independently, i.e. only the underlying mirror, only the top mirror, or on demand both simultaneously. In the simple case, both mirrors will be addressed by an external (here vertical) pressure simultaneously. This air-Bragg-reflector cavity also allows one to have active material placed on the bottom DBR in contact manner, provided that the terminating layer features an adjusted thickness, or incorporated in suspended form. This is a unique approach as far as spectrally-tunable microcavities are concerned.

The refractive index of the IP-DIP photoresist in the visible spectral range is 1.52 at room temperature which exhibits minor changes at low temperature [46]. This allows one to obtain a reasonable refractive index contrast of 1.52/1 between the two materials of the dielectric mirror. The refractive index contrast is not very high, but compared to typical III/V DBRs made of GaAs/AlAs with index contrast in the range of 3.5/3, the dielectric mirror with air and polymer still yields an improvement. Nonetheless, with a large number of mirror pairs, an overall high reflectivity is achievable.

For the TMM calculations, considering the active material to be thin layered semiconducting materials (TMDCs), the design wavelength of the air-Bragg reflector is targeted to be in the range of 600 nm to 800 nm (for the most popular TMDCs with their A-excitons in that range). Note that longer wavelengths, e.g. for near-infrared to infrared intralayer or interlayer excitonic species in TMDC monolayers or 2D heterostructures, respectively, provide more favorable printing conditions than for the here chosen $WS_2$ A-exciton resonance. For detailed explanations regarding the TMM, we refer to the supplementary information of our previous work by Wall et al. [42].

**Figure 2a-d** shows the stopband of an air-Bragg reflector with different layer pair numbers, for four different layer thickness configurations, considering the design wavelength ($\lambda$) to be around 620 nm in air (2.0 eV). The air-Bragg reflector with $\lambda/4$ layer thickness exhibits a reflectivity close to 1 over a 100 nm range when composed of 8 mirror pairs (of air and IP-DIP). The reflectivity as a function of the layer thickness does not change significantly, but it expectedly relies on the number of pairs in the air-

Bragg (**Fig. 2e**). A large number of mirror pairs typically results in a high reflectivity, as evidenced in the calculated reflectivity spectra for all four configurations.

In a fully air-Bragg-based cavity system, the bottom reflector consists of 7 or 8 pairs and the top one of 6 pairs for better out-coupling. The air-Braggs with 7 and 6 pairs possess maximum reflectivities of 98.5 % and 95%, respectively (**Fig. 2e**).

Next, the influence of the layer thickness on the stopband width is briefly summarized. It can be clearly seen that the stopband width changes drastically as a function of the layer thickness. The stopband width for the $\lambda/4$, $3\lambda/4$, $5\lambda/4$ and $7\lambda/4$ air-Bragg reflector amounts to approximately 100 nm, 60 nm, 40 nm, and 25 nm, respectively (**Fig. 2f**).

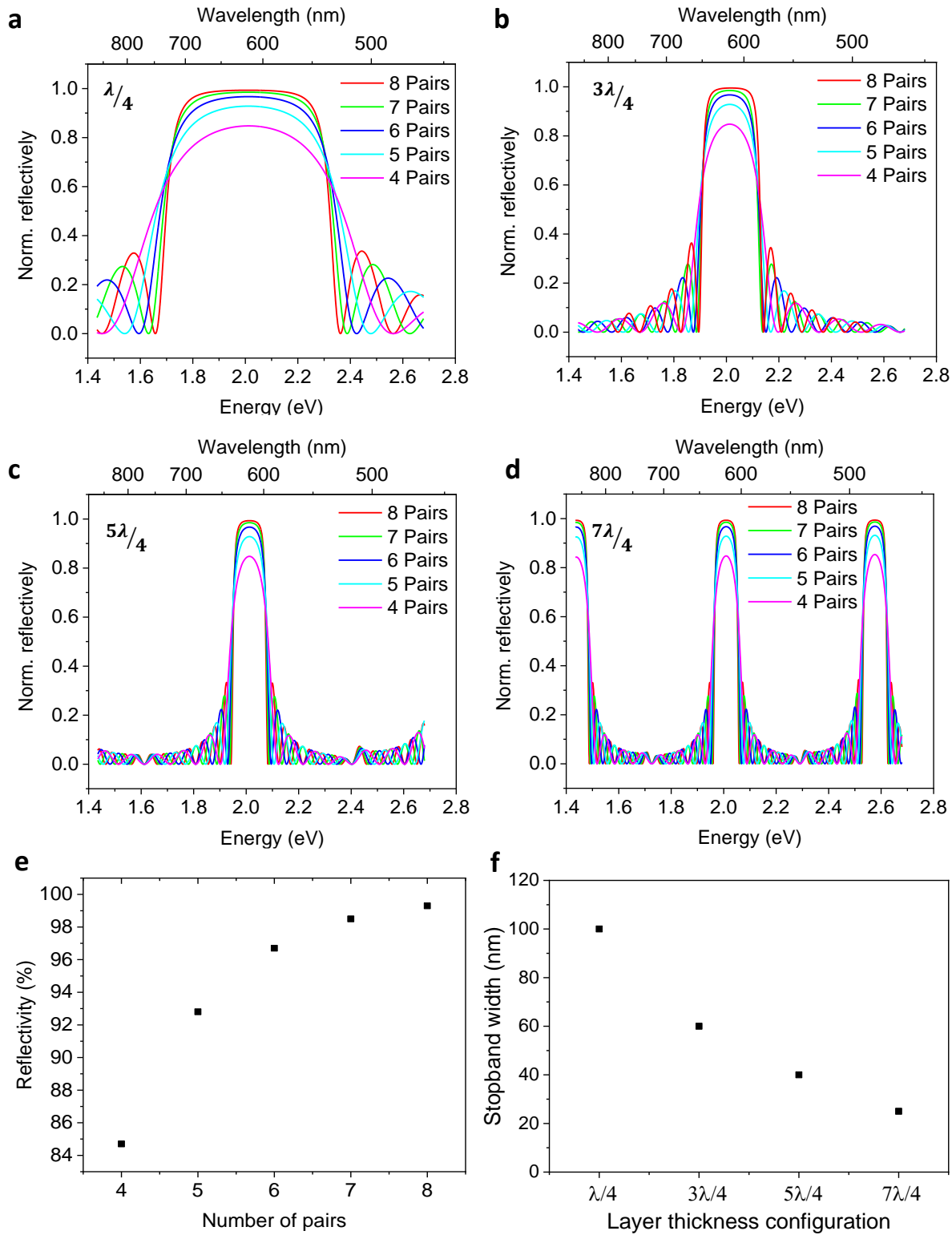

**Figure 2:** *Stopband and reflectivity dependencies of polymer/air-Bragg reflector for different layer thicknesses. Calculated reflectivity spectra for a)* $\lambda/4$ *b)* $3\lambda/4$ *c)* $5\lambda/4$ *and d)* $7\lambda/4$ *material specific (air and polymer) layer thickness. e) Maximum reflectivity given at the Bragg wavelength as a function of the mirror-pair number. Here, the calculated maximum reflectivity exceeds 99% for 8 mirror pairs. f) Extracted stopband width (for 8 pairs) as a function of the layer thickness. Accordingly for a* $\lambda/4$ *layer thickness, the total stopband width is 100 nm, whereas for* $7\lambda/4$ *layers it becomes 25 nm.*

**Stress Analysis of Tunable Polymer/air-Bragg Microcavities**

The crucial stress analysis based on FEA was performed on aforementioned polymer/air-Bragg reflector designs, using a computer-aided design (CAD) tool (see methods section). Here, *Autodesk Inventor* allows simulating the practical pressure-affected cavity configurations, which employ air-Bragg reflectors towards their applications in tunable open-microresonator devices. This is possible due to material-specific mechanical properties allocated to the structure's CAD model. The physical and mechanical properties of the photoresist IP-DIP are summarized in **Tab. 1**.

*Table 1. Key physical and mechanical properties of IP-DIP resin. (If not indicated by \*asterisks, the values are from the Nanoscribe GmbH [47]) \*Ref. [48]\*\*Ref. [49].*

| Resin | Density (liq) (g/cm$^3$) | Density (s) (g/cm$^3$) | Young's modulus (GPa) | Hardness (MPa) | Poisson's ratio | Refractive index |
|-------|--------------------------|------------------------|-----------------------|----------------|-----------------|------------------|
| IP-DIP | 1.14 - 1.19 | 1.2* | 4.5 | 152 | 0.35** | 1.52 |

The CAD of air-Braggs for various layer thickness configurations is shown in **Fig. 3** along with the stress analysis simulation. The air-Bragg reflectors consist of 8 layers with $i$ = 1,3,5,7 quarter-wavelength layer thickness (**Fig. 3a, b, c, d,** respectively) and cover each an area of 50 x 50 µm$^2$. The gradual increment or decrement in the applied pressure leads to the deformation of the polymer parts of the air-Bragg structures. The pressure is solely applied on the (here two opposing) side pillars/bars which mechanically support the quasi-free-standing sub-micrometer-thick polymer layers in the vertical microstructure. The upper surface area of each pillar is 20 x 50 µm$^2$.

To fulfill their purpose, the deformation of these pillars causes a noticeable change in polymer layer separation, which gives the overall microresonator platform the intended functionality to tune/detune cavity modes by deforming the air-Bragg structure. The application of external pressure on such a small structure modifies the thickness of all air layers considerably while leaving the polymer layers basically unchanged. Thereby, the structural modification results in the desired shift of the reflectivity spectrum and ultimately the spectral position of the optical cavity modes in a complete microresonator. However, the tunability of the air-Bragg reflector is limited, on the one hand, by the pressure-bearing capacity (i.e. maximum compression and expansion) of the photoresist IP-DIP, and on the other hand, by the pressure degree, at which the deformation still provides a functioning DBR based on the pressure-affected effective layer thickness ratio between air and polymer layers.

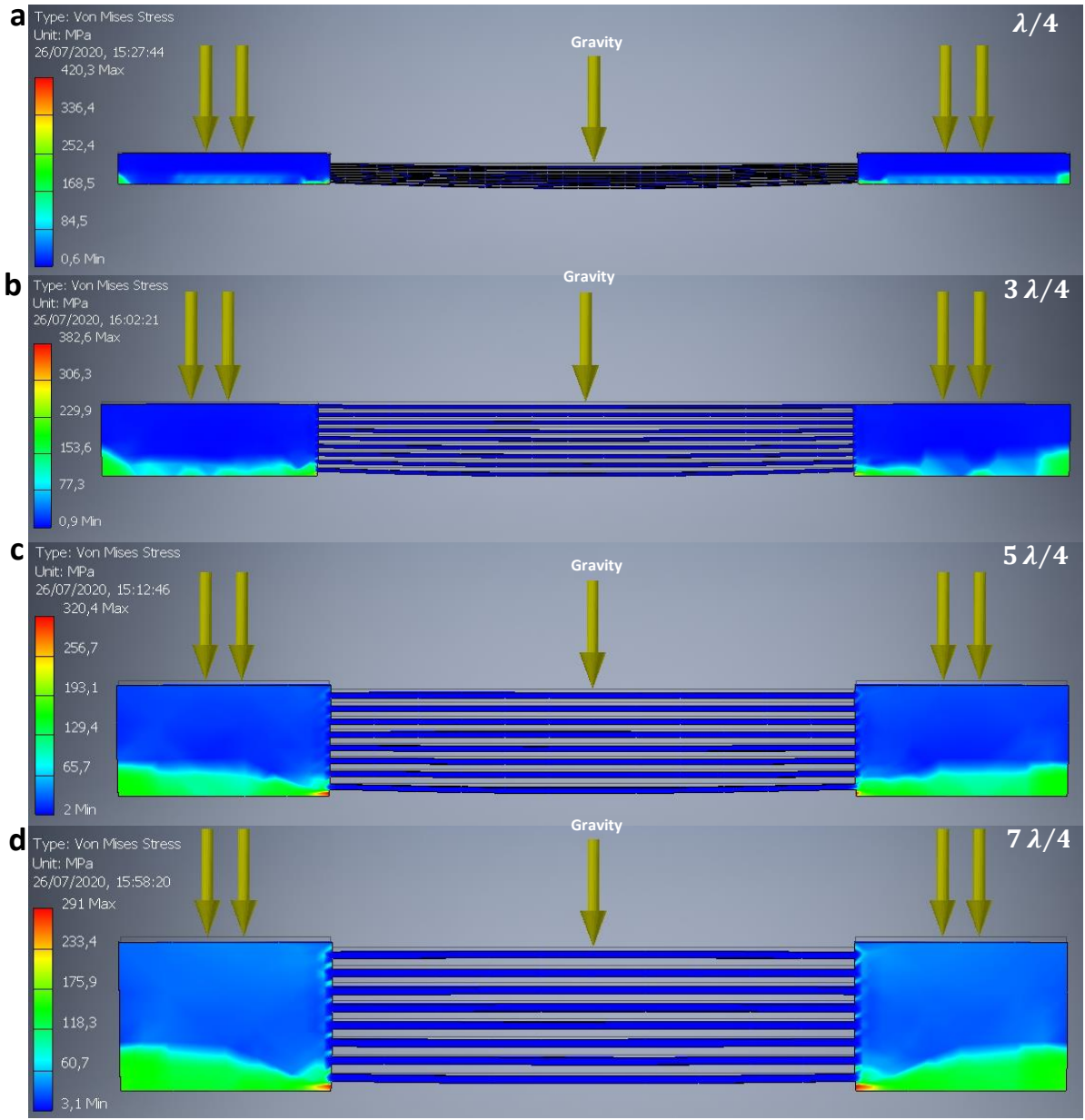

**Figure 3:** *Finite element analysis (FEA) of the polymer/air-Bragg reflector with different layer thickness configurations. FEA simulations for air-Braggs with **a)** $\lambda/4$ **b)** $3\lambda/4$ **c)** $5\lambda/4$ and **d)** $7\lambda/4$ layer thickness demonstrate a gradual deformation of that structure upon application of a uniform pressure of 100 MPa on the side bars of the structure (double arrows). The central arrow indicates the force of gravity applied to the overall structure. The area shaded blue experiences less induced deformation and less pressure (densification) compared to the green and particularly the red colored areas. Here, the floating layers remain unaltered. Note that the thin black lines in the background indicate the initial shape of the structure in the absence of external pressure. In fact, gravity leads to the pronounced bending of the thinnest and, thus, least rigid layers.*

Our simulation-based mechanical analysis indicates that the $\lambda/4$ air-Bragg structure is unstable as can be evidenced by the strong bending and bunching of the photoresist layers. Besides, the air-Bragg with layer thickness $3\lambda/4$ (**Fig. 3b**), $5\lambda/4$ (**Fig. 3c**) and $7\lambda/4$ (**Fig. 3d**) exhibit a clearly more stable behavior under influence of gravity and external pressure due to the improved rigidity. As can be deduced from TMM calculations based on extracted structural information, the effective deformation (in terms of thickness reduction) of the air layers in air-Bragg structures influences the overall stopband of the microcavity. The stopband of the individual air-Bragg reflector experiences an increasing blue-shift in its spectral position when the external pressure is gradually increased as applied on the upper surface of the structure (indicated by double arrows in **Fig. 3**).

The stress analysis method used in this work can also be utilized to study the behavior of the cavity structures upon application of mechanical pressure. To address the targeted effect of cavity-mode tunability by external pressure, two different examples of (printable) cavity configurations under a variation of the applied pressure are discussed, that are (a) the air-Bragg/DBR microcavity and (b) the air-Bragg/air-Bragg microcavity (see **Fig. 1a** and **b**, respectively).

To begin with, **Fig. 4a** demonstrates the tunability induced by applied pressure in the range of 0 to 50 MPa (corresponding to a uniform force of maximally 0.1 N on the two side bars) on the $5\lambda/4$ air-Bragg/DBR microcavity which causes the quality factor $Q = E/\Delta E$ (that is mode energy over linewidth, i.e. full-width-at-half-maximum FWHM) to change drastically. The cavity length of 2.6 µm, whereas $\lambda$ being 620 nm, allows us to determine the cavity mode (C) q = 8. The cavity mode q = 8 clearly changes its spectral position upon application of pressure (**Fig. 4b**), which compresses the overall multilayered air-Bragg structure. In the waterfall diagram of **Fig. 4b**, the TMM calculated reflectivities for the microcavity with arbitrary increment of pressure in steps of 10 MPa are displayed. The resonance of the cavity mode was adjusted to be resonant with the 617 nm A-exciton mode in $WS_2$ at room temperature by fine-tuning the cavity spacer thickness and performing the TMM calculations for this configuration in the absence of external pressure. This ensures one to obtain directly resonance conditions at elevated temperatures around 300 K and also reach a red photon-exciton detuning at low temperature, which is beneficial for pressure-based cavity mode tuning in strong-coupling measurements in cryogenic experiments. An example of the calculated strong-coupling situation in a $WS_2$-air-Bragg microcavity for different pressure levels is displayed in **Fig. 5**.

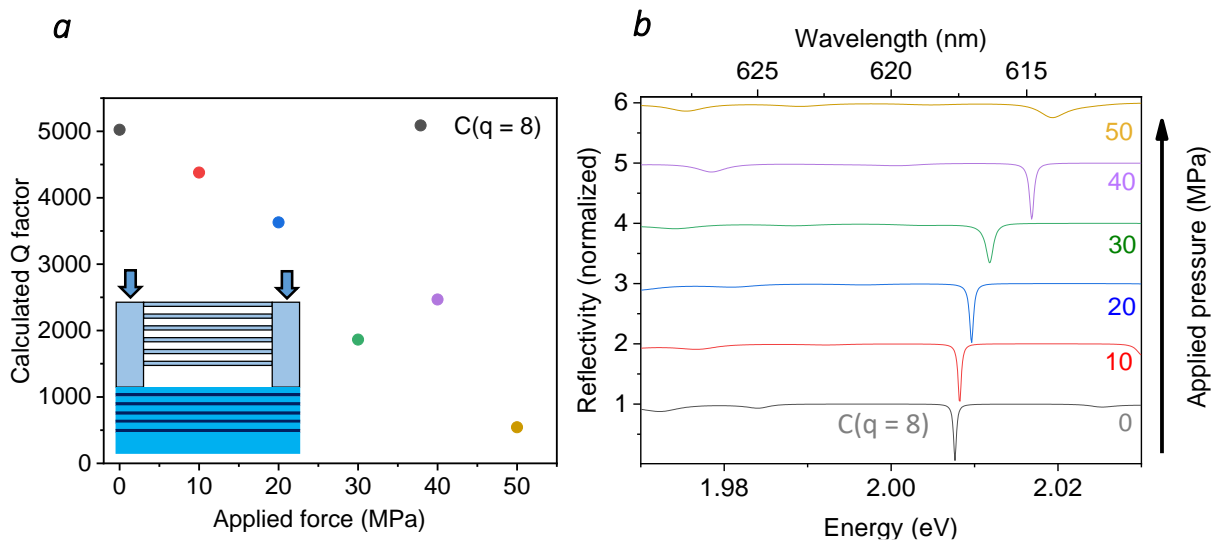

**Figure 4: a)** *Theoretical Q factor of the cavity mode (q = 8) as a function of applied pressure based on calculated reflectivity spectra shown in **(b)** as waterfall diagram with constant vertical offset of 1. The thickness reduction of air layers in the polymer/air-Bragg structures alters the spectral position and Q factor of the cavity mode q = 8 due to the material structure deformation obtained through the uniformly applied pressure.*

The theoretical reflectivity spectrum for each pressure setting is obtained by manually reading out the positions of the individual layers of the (compressed) structure from the FEA simulation results. Thus, the altered air-Bragg structure (with pressure-dependent layer thicknesses) in 1D representation can be fed into the TMM model. Accordingly, the Q factors of the mode q = 8 at 30 and 40 MPa exhibit an unnatural trend, which results from the read-out uncertainties for the respective underlying layer structure visually delivered by the FEA simulation. On the other hand, the expected blue-shift of the mode is attributed to the compressed structure with reduced air-gap thicknesses which is seemingly sufficient in order to obtain the tuning effect. Nonetheless, owing to the increased thickness unproportionality in the polymer/air configuration for increased pressure levels, the resonance conditions in the cavity suffer and a gradually reduced Q factor is obtained.

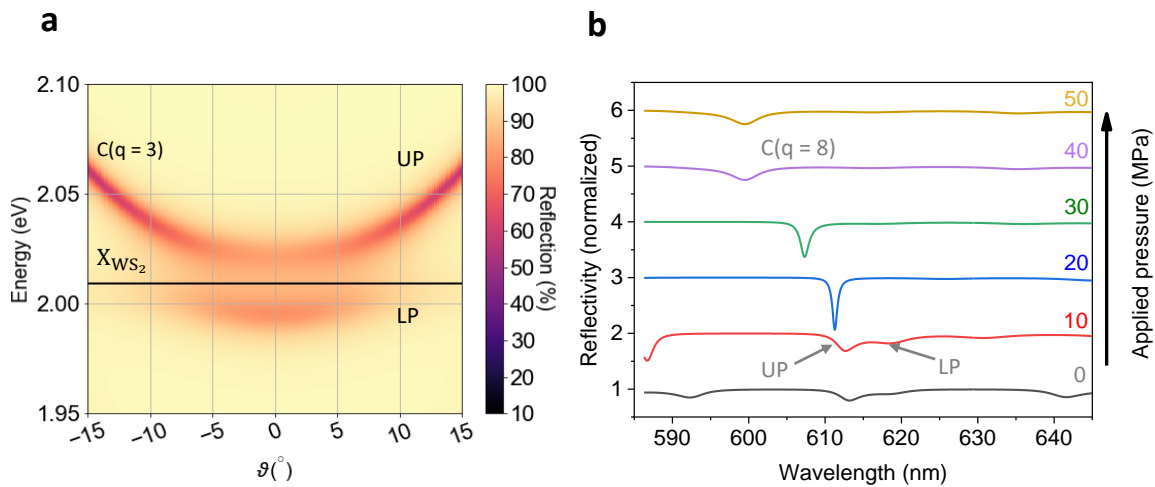

***Figure 5.*** *Calculated angled-resolved reflection for air-Bragg/$WS_2$/planar-DBR microcavity.* ***a)*** *Calculated angle-resolved reflection spectrum (false-color contour diagram) for the $3\lambda/4$ air-Bragg/DBR microcavity illustrating a Rabi splitting of 29 meV obtained with a cavity length of 0.77 µm. Upper (UP) and lower polariton (LP) branches labeled with exciton resonance (black horizontal line) of $WS_2$ at 2.01 eV featuring a hypothetical linewidth of 30 meV, for the bare cavity mode (q = 3).* ***b)*** *Calculated reflection spectra at various exciton-photon resonance detunings demonstrating tunable control over light-matter coupling with clear anticrossing behaviour. The mode detuning is an effect of the gradual application of pressure upon the microcavity structure. In the model, the WS2 monolayer is directly placed on the dielectric mirror's top facet, on which the air-Bragg including a cavity spacer is placed (see* **Fig. 1a**)*.*

In principle, the $SiO_2/Ti_3O_5$ DBR stopband width (150 nm) is much larger than that of the $5\lambda/4$ air-Bragg (40 nm), which allows one to practically shift the stopband of the air-Bragg over a wider spectral region by external pressure. However, a large structural compression may lead to a relatively strong deformation of the air-Bragg configuration from the actual design, which influences not only the spectral position of the cavity mode and stopband, but also the Q factor of the cavity mode (as indicated in **Fig. 4**).

The second approach pursued utilizes a fully air-Bragg-based microcavity design (**Fig. 1b**), and, in this study, the homogeneous external pressure is applied on the whole cavity structure (on its side bars). This particular scenario provides the unique opportunity to investigate light-matter coupling scenarios due to the flexibility with respect to mode detunings, i.e. the energy difference between cavity and emitter modes, in a substrate-independent fashion. Moreover, it can be used to insert suspended 2D-material sheets with support structures, if appropriately designed, as an additional tool to tweak the LMI by changing the position of the ML with respect to the standing electromagnetic (EM) fields inside the cavity. As a consequence of suspension, also the substrate-induced impurities and all other consequent effects get automatically eliminated.

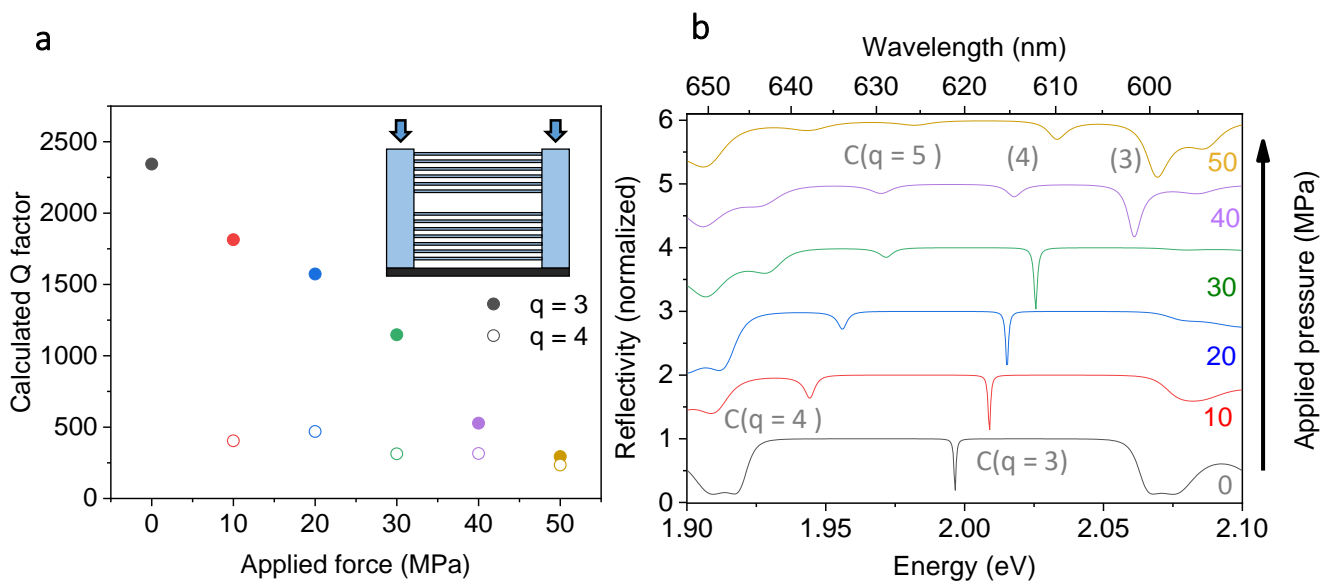

**Figure 6: a)** The influence of the applied pressure (indicated by blue arrows) on the calculated Q factor with schematic of the polymer/air Bragg-reflector microcavity design (inset). **b)** Plot of calculated reflectivity spectra with constant vertical offset of 1, showing the tunability of the polymer/air Bragg-reflector microcavity when a homogeneous pressure is applied (varied between 0 and 50 MPa).

In this configuration, the cavity spacer is similar to the first example proportionally reduced together with air gaps in the Bragg sections. Again, a gradual reduction of the Q factor is obtained (**Fig. 6a**) and the overall cavity stopband experiences a comparable shift. **Figure 6b** demonstrates the tuning of modes in the calculated reflectivity spectra of the air-Bragg microcavity system upon application of external pressure ranging between 0 and 50 MPa. The step-wise reduction of air layer thicknesses causes a controllable blue-shift of the cavity mode's spectral positions (q = 3 and 4), which is accompanied by a reduction of the Q factor with increasing pressure level.

The crucial stress analysis of the air-Bragg structures for tunability of cavity modes with sufficiently high Q factors let us conclude that the targeted strong light-matter coupling scenarios can be theoretically realized. Incorporating quantum dots, florescent dyes or even TMDC monolayers in air-Bragg/DBR and all-air-Bragg microcavities can open the path to flexible resonator-emitter systems for various LMI experiments. Additionally, this approach delivers in-situ control over LMI by an applied vertical force on the surface of the device (i.e. mechanical pressure), which provides the necessary mode tunability by compression of the microcavity structures. Inspired by these theoretical results, work is ongoing towards the optimization of the printing parameters for such air-Bragg structures to obtain the overall optical quality and mechanical stability along with the investigation of various approaches to systematically induce the necessary thickness-reducing deformations in polymer/air-layers in a controlled and controllable fashion for larger and, foremost, predictable tunability of high-quality resonator modes.

**Conclusion**

To summarize, two spectrally-tunable microcavity configurations based on polymer/air Bragg reflectors (air-Braggs) with accessible open cavity spacer were discussed for weak and strong LMI, such as with various types of nanomaterials, molecules as well as quantum dots. Building upon the capability of 3D nanoprinting these polymer/air layered structures with sub-micrometer precision, this study explored the concept of mechanical-pressure-induced cavity mode shifts through air-Bragg structure compression. Our modelling work indicates that by using state-of-the-art technology, one can in principle realize optical microcavity systems on demand for various spectral regions provided that the target wavelength is compatible with the printing precision and minimal feature size (voxel size) of the 3D nanoprinting technique. The crucial stress analysis based on FEA simulations reveals the tunable nature of the polymer/air Bragg structures upon pressure application, leading to a controllable energy shift of the microresonator modes. Such versatile light-matter interface that can be conveniently deposited on different surfaces at the position of interest consist at minimum of one such "air-Bragg" reflector and can for the discussed designs theoretically exhibit Q-factors up to 5000. This implies that different light-matter coupling scenarios with flexibility in the cavity-emitter resonance detuning, even after production, can be realized with the help of such polymer/air-based optical microcavities.

In the future, precise laser nanoprinting of (integrated) photonic structures directly on the chip will be a common practice, and an approach to achieve optical microreflectors and microresonators such as the here presented one can be very useful in many specific scenarios, including for LEDs and photovoltaics, nanolasers and nonclassical light sources, optical filters or couplers and nonlinear optical elements, as well as optical sensing through LMI.

**Methods**

**Stress analysis of CAD structures:** The stress analysis function provided by *Autodesk Inventor Professional 2019* was used to determine the impact of vertical external pressure (force per area) on different polymer/air-Bragg structures using CAD models. The stress analysis function based on a finite element analysis (FEA) allows one to assign the applied force on the surface of the CAD model and to obtain mechanical deformation for every setting of externally applied force (represented by the double arrow in the shown figure) and under influence of gravity (single arrow). The hereby obtained prediction of the overall structural response was later used to model reflectivity spectra (using manually extracted position changes of the layered structure for given pressure level).

**Calculation of spectra:** The targeted microreflector and microresonator systems were modeled using the transfer matrix method (TMM) regarding their standing-wave light-field profile and reflectivity spectra. Based on theoretical data and considerations with chosen design parameters, reflectivity spectra towards strong light-matter coupling with a virtual 2D semiconductor inside the cavity were equally obtained. The optical properties such as reflection, transmission and angle-resolved spectra of multi-layered open cavity structures were calculated using the TMM-based simulation code as used in Wall et. al [42].

**Acknowledgement**

The authors would like to thank F. Wall for providing her optimized TMM frame for 2D-materials cavity simulations, and Dr. Henning for helpful discussions and insights on 3D nanoprinting. Access to the clean room of the MiNa Lab at Justus Liebig University (JLU), Giessen, through the subscription of the Semiconductor Photonics Group, Marburg, is acknowledged. The authors are furthermore grateful to the (former) members of their 2D Materials team, Marburg, for fruitful discussions.

**Author contributions**

ARI initiated the study and conceived the concepts for printed tunable microcavities for LMI with 2D materials and nanoparticles. CCP performed the theoretical optical analysis using the TMM and simulated mechanical properties. CCP and ARI designed the structures, outlined the simulations and evaluated the data. The results were summarized in a manuscript by both authors.

**Corresponding author**

a.r-i@physik.uni-marburg.de

**Authors' statement/Competing interests**

The authors declare no conflict of interest

**Additional information**

Supplementary Information accompanies this paper

**Supplementary Information**

## 1. The air-Bragg-reflector microcavity with inserted 2D membrane

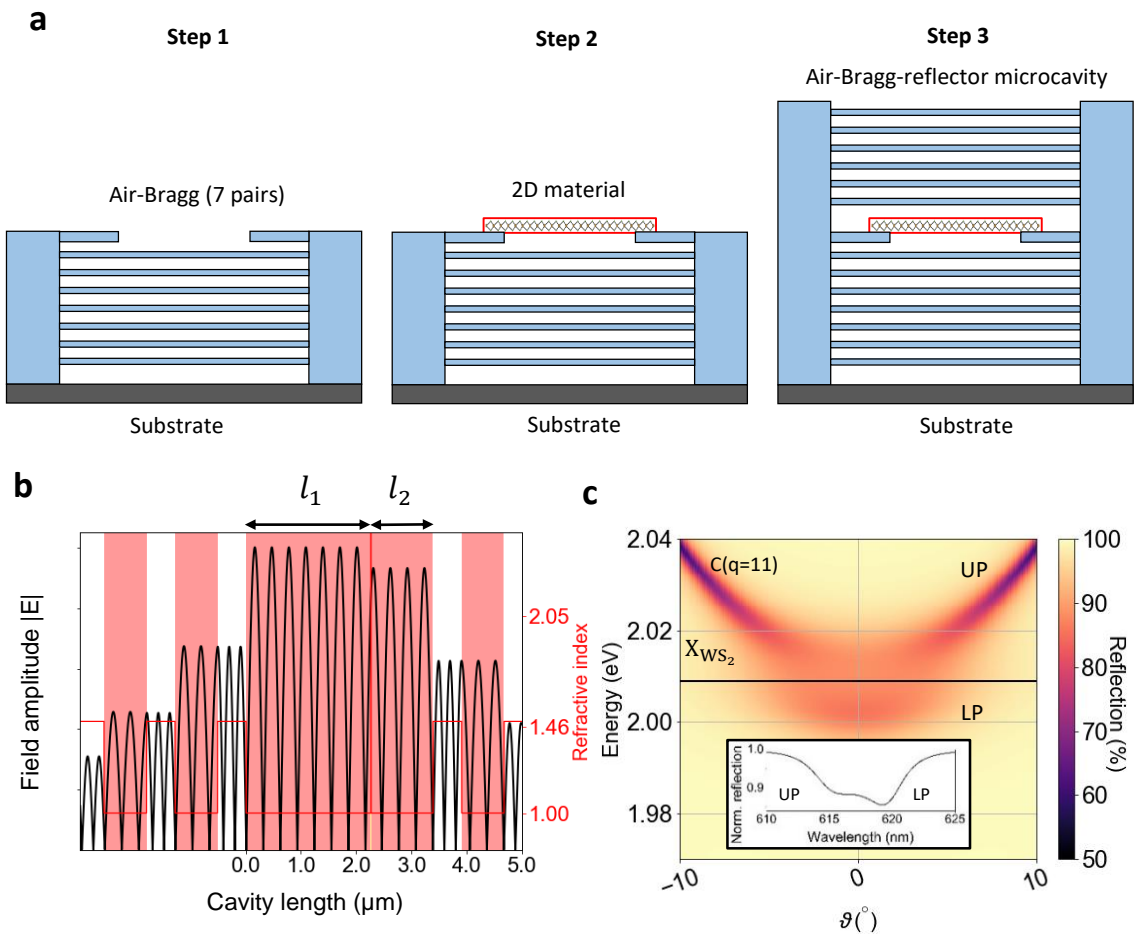

**Figure S2. Numerical calculations for an all-air-Bragg-reflector microcavity.** *a) Schematic of air-Bragg-reflector microcavity with suspended $WS_2$ monolayer (red color slab), indicating the step-wise assembly from left to right in three steps. b) Field distribution inside the air-Braggs microcavity with $WS_2$ ML at field maximum. $l_1$ and $l_2$ are the separation between monolayer and the upper and lower air-Bragg reflector, respectively. c) Calculated angle-resolved reflection (false-color contour plot) of the closed cavity consisting of air-Bragg reflectors with $5\lambda/4$ layer thickness which exhibits a Rabi splitting of $13 \pm 3$ meV at a total cavity length of 3.4 µm. The upper (UP) and lower polariton (LP) branches are labeled with the exciton resonance of $WS_2$ at 2.01 eV (black horizontal line), for the bare cavity mode (q = 11). Inset: line spectrum of the microcavity reflection at normal incidence angle, corresponding to almost zero detuning between exciton and photon mode (spectral resonance).*